\documentclass[]{aa} 

\newcommand\nice[1]{#1}    \newcommand\subm[1]{}   

\newcommand\dosingle[1]{#1}   \newcommand\dodouble[1]{}  
\subm{ \renewcommand\dosingle[1]{ }   \renewcommand\dodouble[1]{#1} }

\usepackage{graphicx}  

\usepackage{txfonts}

\def\SSS{Sect.~}
\newcommand\asr{AdvSpaceRes} 

\nice{ \newcommand\mycaptionfont{\protect\footnotesize} }

\def\gtapprox{\,\lower.6ex\hbox{$\buildrel >\over \sim$} \, }
\def\ltapprox{\,\lower.6ex\hbox{$\buildrel <\over \sim$} \, }
\def\propapprox{\,\lower.6ex\hbox{$\buildrel \propto\over \sim$} \, }

\def\arcs{\ifmmode {'' }\else $'' $\fi}     
\def\arcm{\ifmmode {' }\else $' $\fi}       

\def\fr7{7$ \hskip -0.9ex \vrule height0.8ex width0.8ex depth-0.73ex
                                                     \hskip0.1ex$}
\def\frtoday{Le\space\number\day\space\ifcase\month\or
  janvier\or f\'evrier\or mars\or avril\or mai\or juin\or
  juillet\or ao\^ut\or septembre\or octobre\or novembre\or 
d\'ecembre\fi\space \number\year}

\newcommand\joref[5]{#1, #5, {#2, }{#3, } #4}  
\newcommand\confref[5]{#1, #5, {#2, }{#3, } #4} 
 
\newcommand\epref[3]{#1, #3, #2}

\newcommand\cqg{ClassQuantGra}   %
\newcommand\hMpc{\mbox{$h^{-1}$ Mpc}}
\newcommand\hGpc{\mbox{$h^{-1}$ Gpc}}

\newcommand\Omm{\Omega_{\mbox{\rm \small m}}}
\newcommand\Omtot{\Omega_{\mbox{\rm \small tot}}}

\newcommand\Lselec{L_{\mbox{\rm \small selec}}}
\newcommand\npairs{n_{\mbox{\rm \small pairs}}}

\title{Cosmic crystallography using short-lived objects\\
 -- active galactic nuclei}

\author{Andrzej Marecki\inst{1},
Boudewijn F. Roukema\inst{1}
\and Stanislaw Bajtlik\inst{2}
}

\institute{Toru\'n Centre for Astronomy, N. Copernicus University,
ul. Gagarina 11, PL-87-100 Toru\'n, Poland
\and
Copernicus Astronomy Centre, ul. Bartycka 18, PL-00-716 Warsaw, Poland}

\date{Received 6 October 2004/Accepted 5 January 2005}

\titlerunning{Cosmic crystallography using AGNs}
\authorrunning{Marecki, Roukema \& Bajtlik}

\abstract{
Cosmic crystallography is based on
the principle that peaks in the pair separation histogram (PSH) of objects
in a catalogue should be induced by the
high number of topologically lensed pairs that
are separated by Clifford translations, in excess to ``random'' pairs
of objects.
Here we present modifications of this
method that successively improve the signal-to-noise ratio
by removing a
large part of the noise and then false signals induced by
selection effects.
Given the transient nature of the most
readily available tracer objects, active galactic nuclei (AGNs),
the former is possible because 
a natural filter for removing many of the noise pairs is available:
when counting pairs of objects in order to create PSHs,
only those
with nearly identical redshifts need to be counted.
This redshift filter (a maximum value of $\Delta z/z = 0.005$)
was applied to a
compilation of AGN catalogues. Further noise was removed by
applying a second filter,
a maximum angle
$\Delta \theta =0.075$~rad, and a minimum number of pairs
$\protect\npairs=3$ to find each ``bunch of pairs'' (BoP) where the
{\em vectors} (in Euclidean comoving space) defined by pairs
are required to be nearly equal,
whereas in the PSH only the {\em lengths} must be nearly equal.
These filters reveal significant signals, which, however,
are due to selection effects. A third filter,
a minimum length $\Lselec=150$~{\hMpc}
between the (parallel) vectors in a BoP,
is found to effectively remove these selection effect pairs.
After application of
these successive filters,
no significant topological signal was found.
}

\begin{document}

\maketitle

\dodouble{\clearpage} 


\def\fPSHdeltaz{
\begin{figure}[ht]
\centering
\includegraphics[scale=0.53]{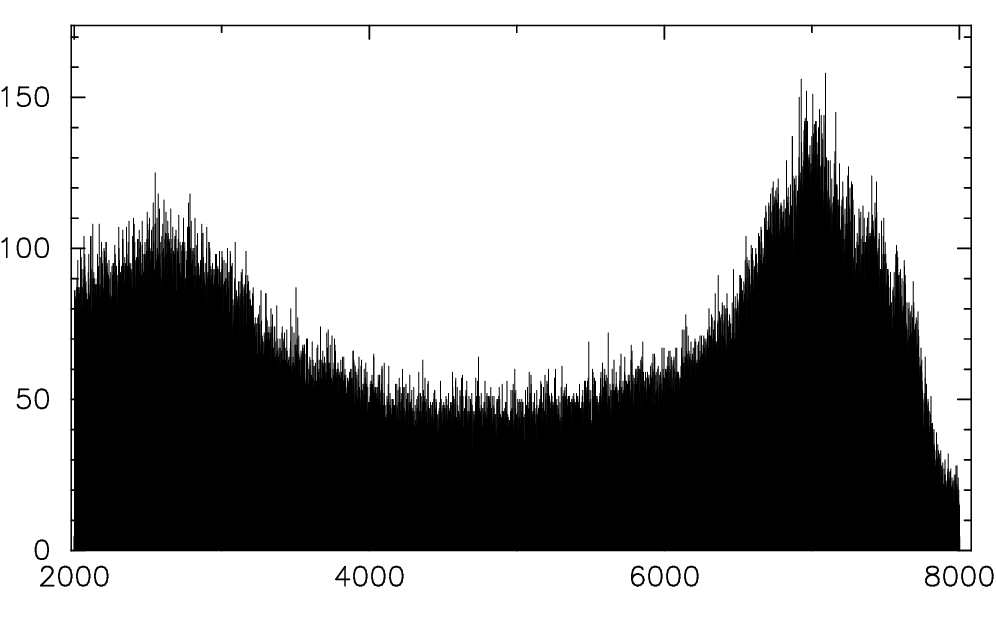}
\caption[]{ \mycaptionfont
PSH (pair separation histogram) 
for the objects in our sample. The only filter applied is 
requiring the redshifts to
both objects in a pair to be very similar --- 
\SSS\protect\ref{s-dzfilter}. The 
error tolerance in the determination of redshifts is 0.5\%
difference in redshifts in a pair. The horizontal axis shows separations of
objects in pairs in units 
of {\hMpc}. The vertical axis shows the number of pairs in
a separation bin equal to 1{\hMpc}.}
\label{f-PSHdeltaz} 
\end{figure}
} 

\def\fspectr{
\begin{figure}[ht]
\centering 
\includegraphics[scale=0.53]{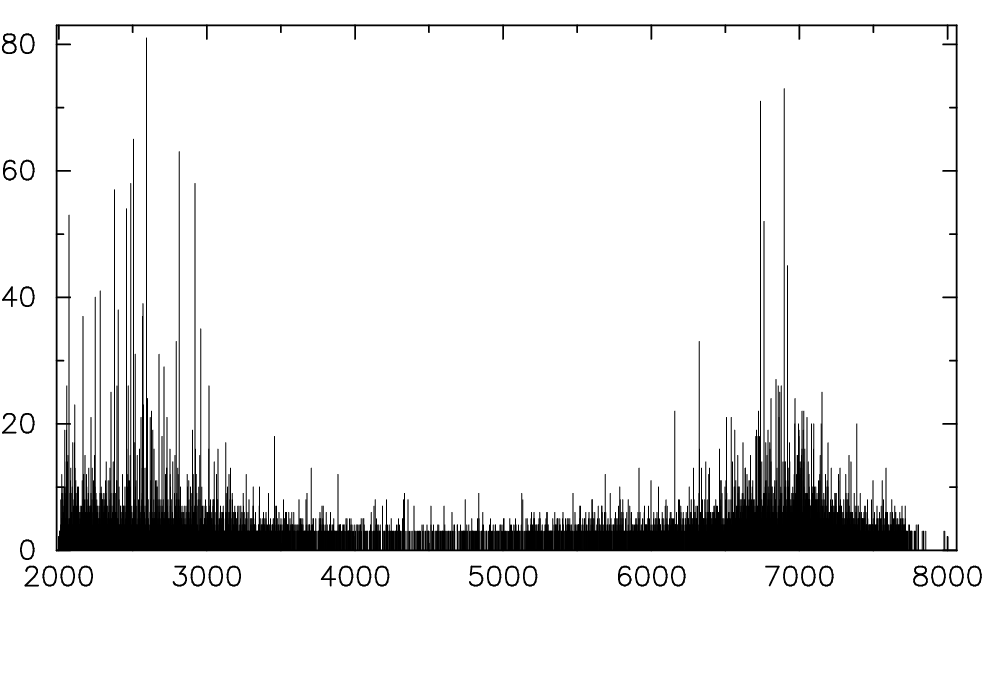}
\caption[]{ \mycaptionfont
PSH of Fig.~\protect\ref{f-PSHdeltaz} modified according to the BoP
filter (\SSS\ref{s-BoPfilter}). The horizontal axis shows separations of
objects in pairs in {\hMpc}. The vertical axis shows the number of nearly
parallel pairs in a bunch. As in the case shown in 
Fig.~\protect\ref{f-PSHdeltaz} we
allowed for 0.5\% difference of redshifts in each pair. The angular
separations of pairs in a bunch of pairs is $\Delta\theta < 0.075$~rad, 
and $\npairs=3$.}
\label{f-spectr}
\end{figure}
} 

\def\fpolar{
\begin{figure}[ht]
\centering 
\includegraphics[scale=0.53]{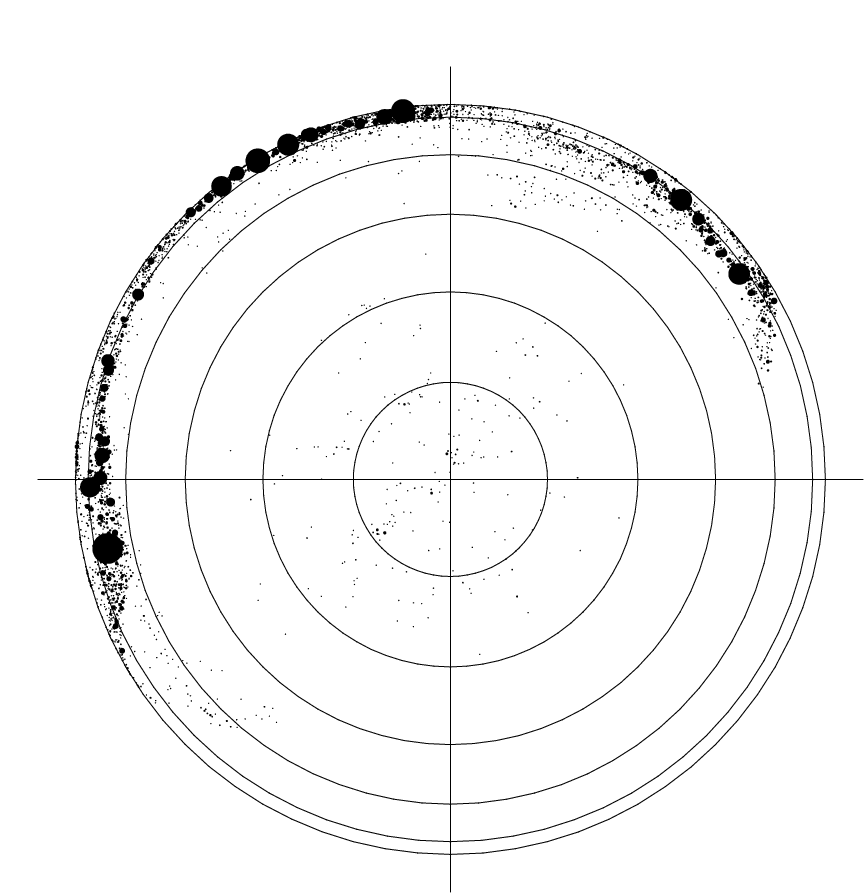}
\includegraphics[scale=0.53]{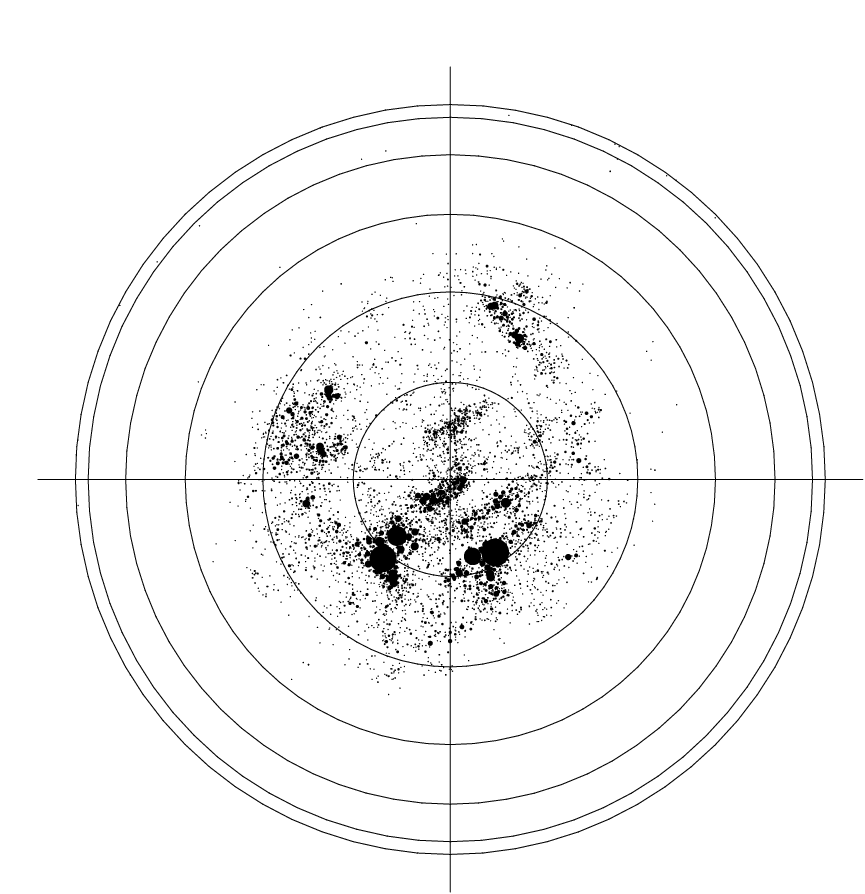}
\caption[]{ \mycaptionfont
The distribution of directions the BoPs 
(\SSS\ref{s-BoPfilter})
point to in galactic coordinates.
The plots are in polar coordinates: the north galactic pole is in the 
centre, circles are plotted for $b=75\degr,60\degr,45\degr,30\degr, 
15\degr,0\degr$. The upper panel is 
for pairs shorter than 5{\hGpc}; the lower panel is 
for pairs longer than 5{\hGpc}.
The radii of filled circles marking the positions are proportional
to the number of pairs in a particular BoP.}
\label{f-polar}
\end{figure}
} 

\def\fclustp{
\begin{figure}[ht]
\centering 
\includegraphics[scale=0.53]{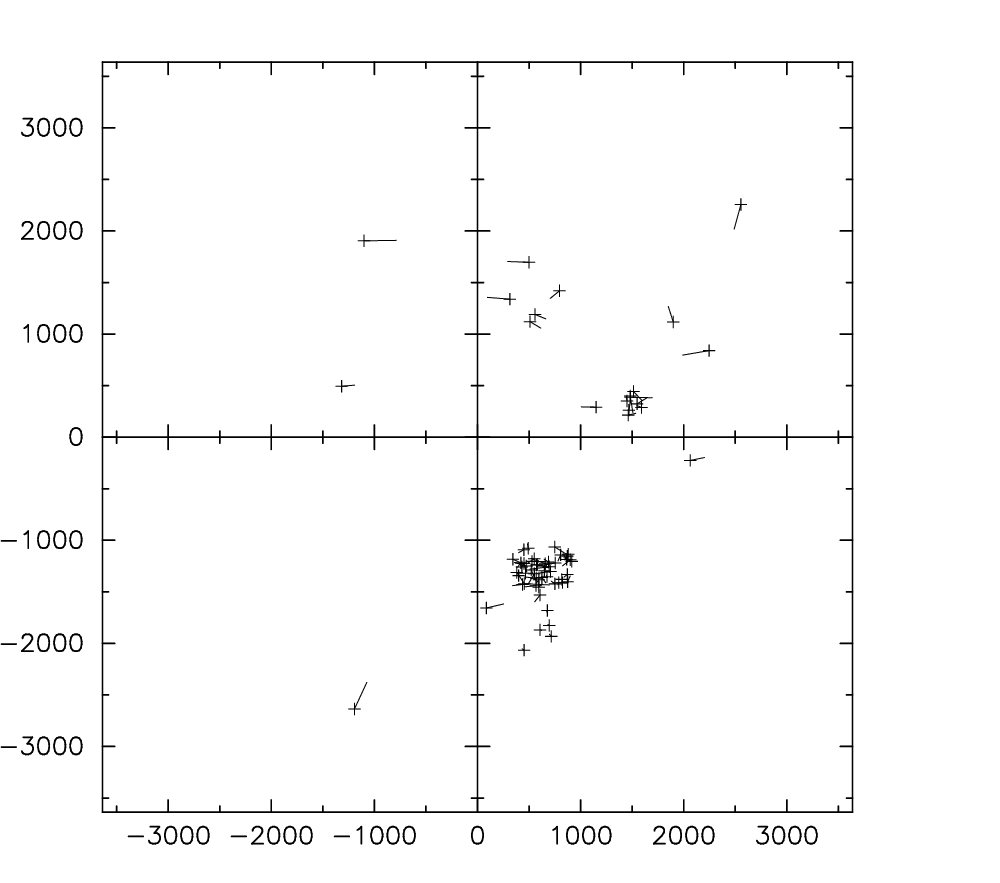}
\includegraphics[scale=0.53]{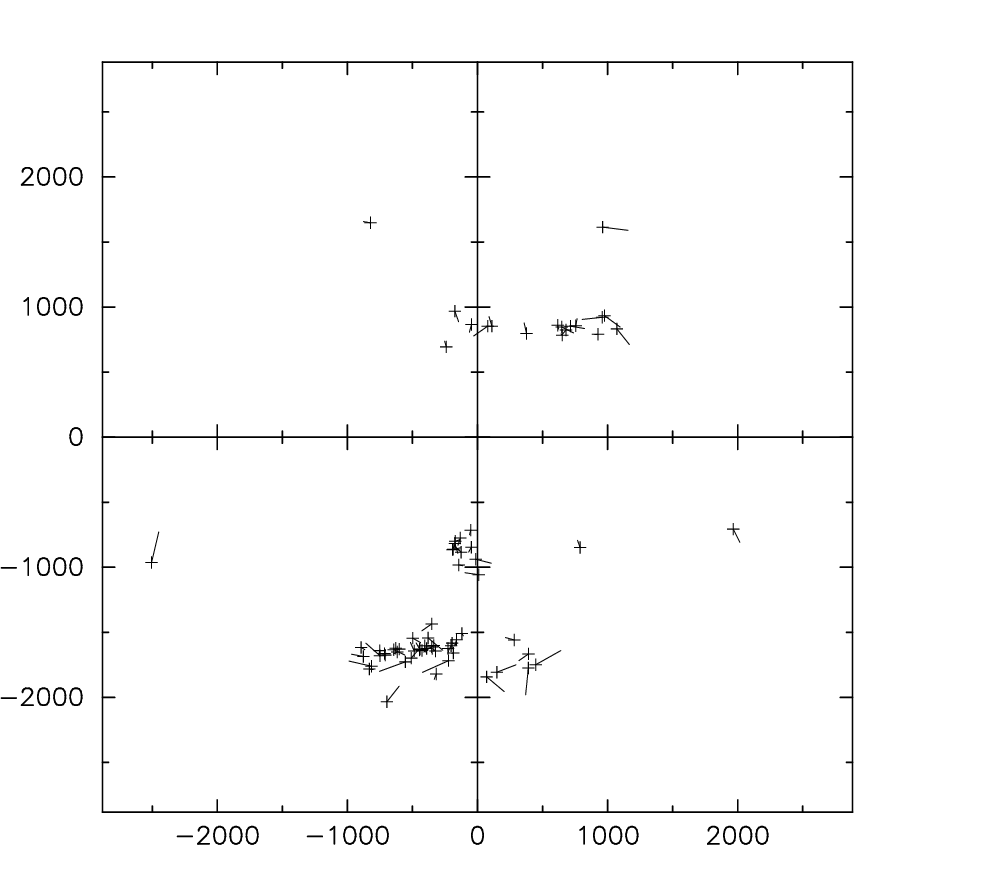}
\caption[]{ \mycaptionfont
Locations of pairs in the BoPs represented by the two highest peaks around
7000{\hMpc} in Fig.~\ref{f-spectr}, shown as their projections onto
the equatorial plane of the 
cartesian coordinate system in which the direction of the
$z$ axis is aligned with the direction the BoP points to. (The
coordinates are in {\hMpc}.) An overwhelming majority of pairs are highly 
concentrated 
in one or two regions, which is what would be expected from selection
effects which mimic a topological signal. A real 
topological isomorphism should generate BoPs throughout
the plane, not just in a few regions.}
\label{f-clustp}
\end{figure}
} 

\def\fLselec{
\begin{figure}[ht]
\centering 
\includegraphics[scale=0.53]{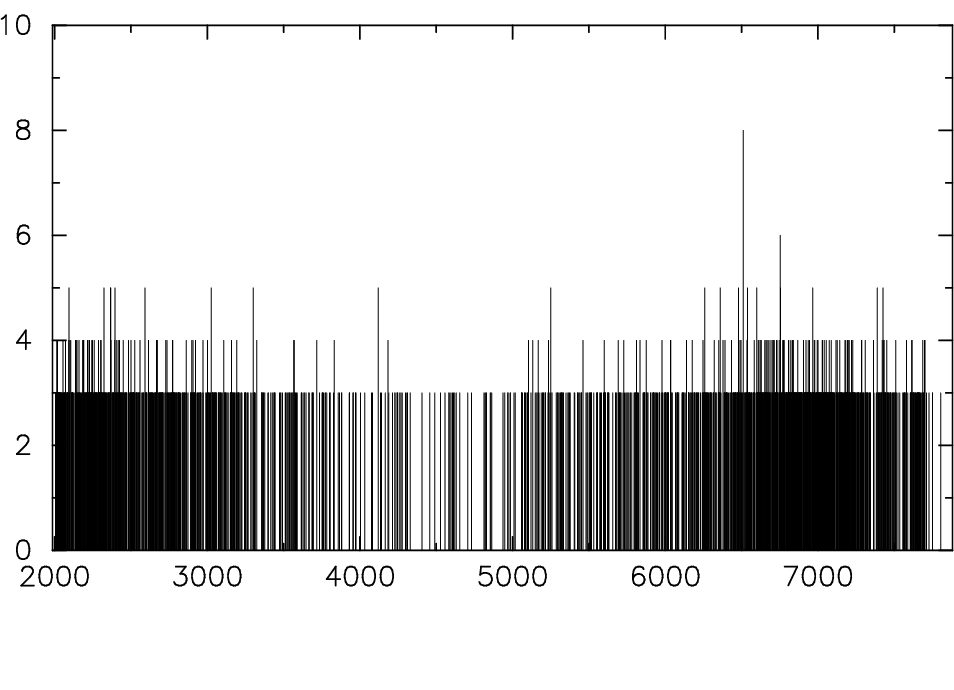}
\caption[]{ \mycaptionfont
PSH from Fig.~\ref{f-spectr} with the $\Lselec=150${\hMpc} criterion imposed.
The horizontal axis shows separations of objects in pairs in $h^{-1}$~Mpc.
The vertical axis shows the number of nearly parallel pairs in a bunch. As in
the case shown in Figs.~\ref{f-PSHdeltaz} and \ref{f-spectr}, we allowed 
for 0.5\% difference of redshifts in each pair; 
and as in Fig.~\protect\ref{f-spectr}, the minimum angular separation of 
pairs in a bunch is $\Delta\theta < 0.075$~rad with a minimum of 
$\npairs=3$ pairs. Clearly, the $\Lselec$ criterion removes most selection
effect BoPs.}
\label{f-PSHLselec}
\end{figure}
} 

\def\ffinal{
\begin{figure}[ht]
\centering 
\includegraphics[scale=0.53]{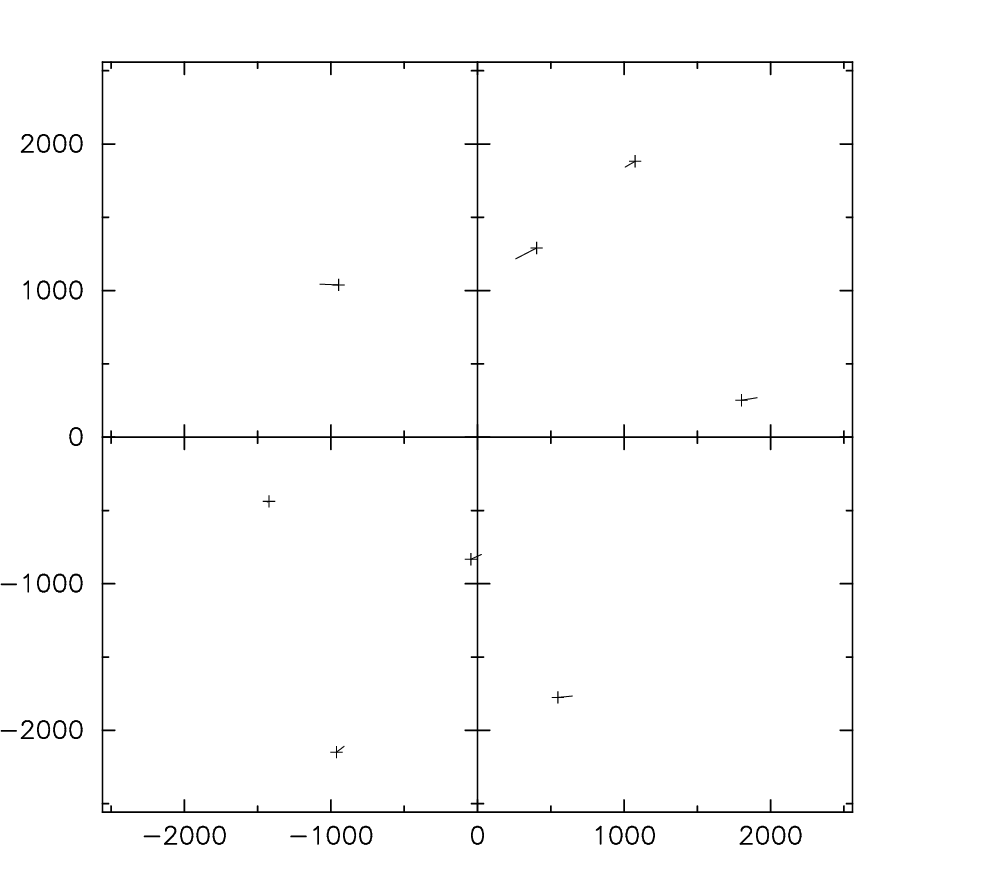}
\caption[]{ \mycaptionfont
Locations of pairs in the BoP represented as the highest peak
in Fig.~\protect\ref{f-PSHLselec} shown in a cartesian coordinate system
analogous to that in Fig.~\protect\ref{f-clustp}.
Thanks to the $\Lselec$ criterion, 
there are no selection-effect-generated clumps of pairs. The remaining
pairs are either the tail of a distribution of BoPs satisfying all the
filters; or they could possibly be a topological signal.}
\label{f-final}
\end{figure}
} 

\newcommand\fdNdz{
\begin{figure}[ht]
\centering 
\includegraphics[scale=0.53]{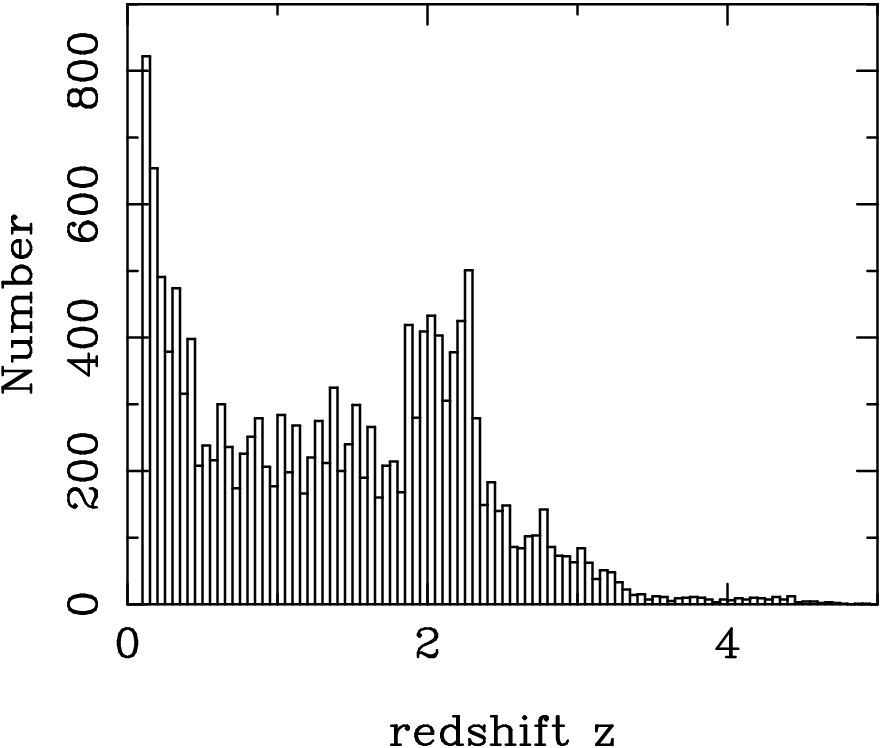}
\caption[]{ \mycaptionfont
Redshift distribution of the sample used for analysis --- see 
\SSS\protect\ref{s-generic}.}
\label{f-dNdz}
\end{figure}
} 

\newcommand\fdNdmag{
\begin{figure}[ht]
\centering 
\includegraphics[scale=0.53]{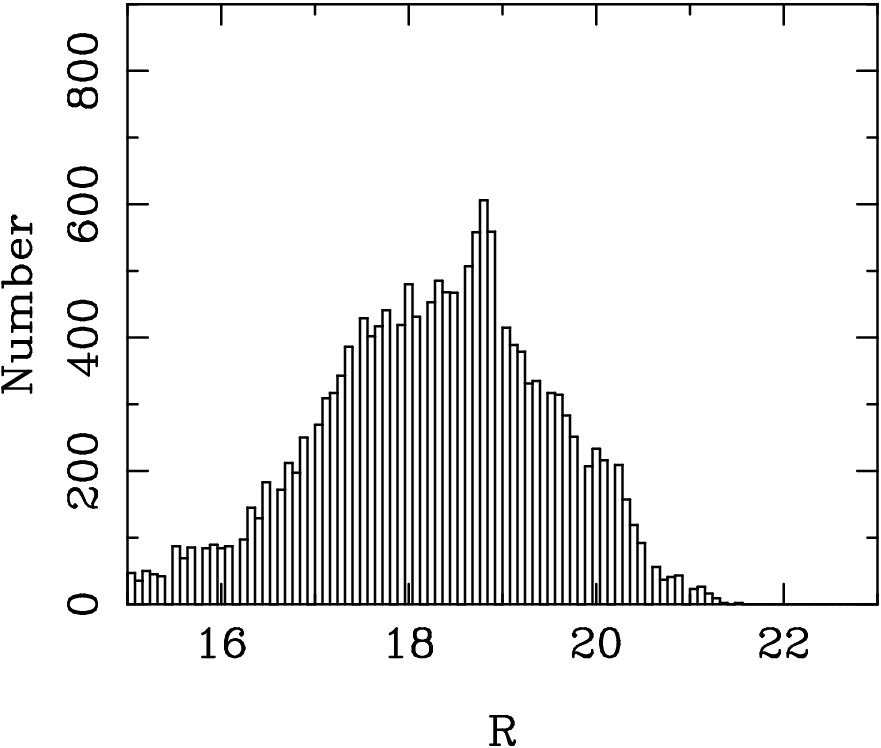}
\includegraphics[scale=0.53]{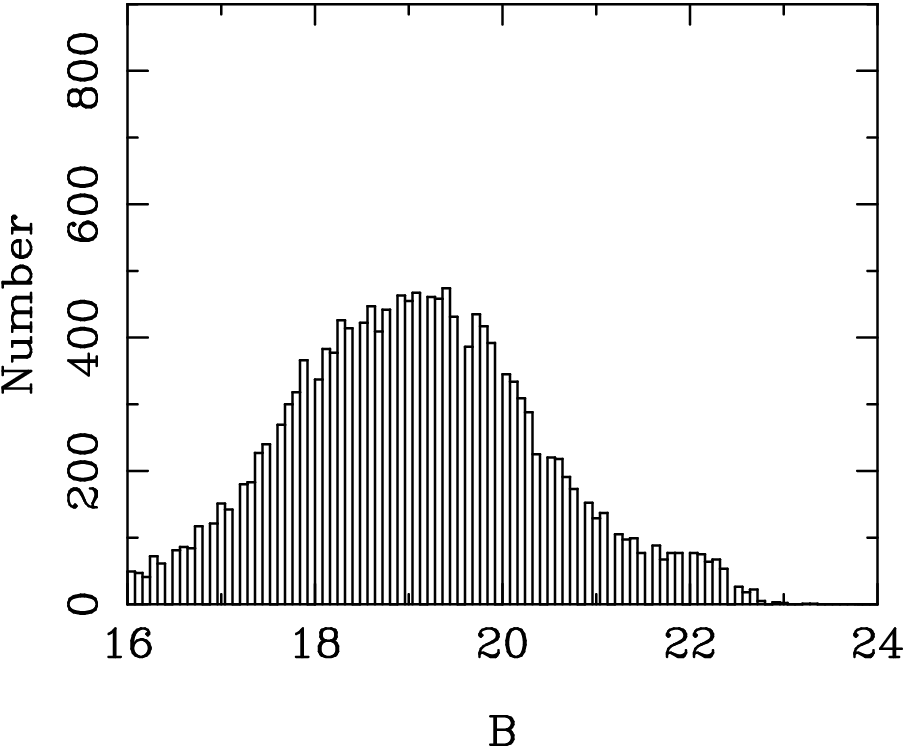}
\caption[]{ \mycaptionfont
Apparent $R$ (upper panel) and $B$ (lower panel) magnitude distribution
of the sample used for analysis --- see \SSS\protect\ref{s-generic}.}
\label{f-dNdmag}
\end{figure}
} 


\section{Introduction}

General relativity (GR) relates the local geometry and content of space-time
in the limit as
one approaches any given point. Together with
assumptions about approximate local homogeneity and local
isotropy\footnote{For comments on ``local'' versus ``global'' homogeneity
and isotropy, see \nocite{Rouk02asr}{Roukema} (2002a).}, the equations of GR relate the metric
properties of space (curvature) and the average material contents and
dynamics (density, expansion rate, cosmological constant) of the Universe.

GR is, however, only a local theory that does not constrain global
geometry: the curvature of space does not tell us whether the Universe
is finite or infinite in spatial volume (or
mass). \nocite{Schw00,Schw98}{Schwarzschild} (1900, 1998) pointed out a century ago that a flat,
homogeneous universe can be finite if the topology of space is not
trivial, i.e. if space is multiply
connected. De Sitter (1917) \nocite{deSitt17}   
and \nocite{Fried23}{Friedmann} (1923) also noted this, while
\nocite{Fried24}{Friedmann} (1924) pointed out that even a universe of negative curvature
could be homogeneous but finite. In contrast to local geometry, which
is constrained by GR, there is neither consensus on a physical theory that
would predict what the topology of the Universe should be nor any
theory that would relate that topology to dynamical properties of the
matter distribution, even though work on some elements of what might
be needed has begun, e.g. \nocite{DowS98,Hawking84a,Hawking84b,ZelG84}{Dowker} \& {Surya} (1998); {Hawking} (1984a, 1984b); {Zeldovich} \& {Grishchuk} (1984).

Nearly all methods for detecting the topology of the Universe rely on
the principle of multiple topological imaging of either discrete objects or
of the thermally radiating plasma seen in the cosmic microwave background
radiation. This phenomenon is called
`topological lensing'. Detection of topological lensing would allow
us to directly study the evolution of individual objects over a cosmological
time-scale, as the same object would be visible at different cosmic epochs. 

Although the amount and the quality of extragalactic observations
available is growing constantly, and many different observational
strategies are possible, detection of topological lensing has so far
proved to be very difficult. Reviews of different strategies include,
e.g. \nocite{ULL99b,LR99,Rouk02topclass,RG04}{Uzan} {et~al.} (1999a); {Luminet} \& {Roukema} (1999); {Roukema} (2002b); {Rebou\c{c}as} \& {Gomero} (2004). 

One of the methods, ``cosmic crystallography'' \nocite{LaLu95,LLL96}({Lachi\`eze-Rey} \& {Luminet} 1995; {Lehoucq} {et~al.} 1996), was
explained and applied to some catalogues of observational data, but
no significant detection has so far been obtained.

Is further application of this method still worth trying? 
Arguments in favour include:
\begin{list}{(\roman{enumi})}{\usecounter{enumi}}
\item The microwave background 
observations of DASI, BOOMERanG, MAXIMA, DMR and
CBI lead to $\Omtot\simeq 1$ \nocite{Sievers03}({Sievers} {et~al.} 2003). This estimate has been
strongly underpinned by the WMAP-based result:
$\Omega_{tot}=1.02\pm0.02$ \nocite{WMAPSpergel}({Spergel} {et~al.} 2003). The fact that the 
Universe inside of the observable 
horizon appears to be nearly flat makes cosmic crystallography a viable
method.
\item Active galactic nuclei (AGNs) have short lifetimes as active (easily
visible) objects, leading to (approximately) removable noise 
in the pair histograms. In this work, a method of removing these pairs and
improving the signal-to-noise ratio is introduced.
\end{list}

The main argument against applying cosmic crystallography is that 
recent attempts at constraints using the WMAP microwave background data
have so far failed to detect matched circles for torus-like 
models \nocite{CSSK03}(e.g. {Cornish} {et~al.} 2004). 
Also, analyses of the WMAP data are accumulating 
that suggest that the global topology
of the Universe may be that of the Poincar\'e dodecahedral space 
\nocite{LumNat03,RLCMB04}({Luminet} {et~al.} 2003; {Roukema} {et~al.} 2004). Clearly, at any time well past the 
quantum epoch, $t \gg 10^{-43}$~s, the comoving spatial section of the 
Universe cannot simultaneously 
be both a Poincar\'e dodecahedral space and a toroidal 3-manifold.

These reasons are not, however, sufficient for choosing not 
to develop and apply several 
tests for such a fundamental property
of the Universe by independent methods, using independent sets of 
objects. 
The cosmic microwave background (CMB) remains difficult to 
interpret and filter for foreground contamination and for the 
integrated Sachs-Wolfe Effect, and even 
the internal linear combination WMAP map (ILC) 
is found to include significant foreground components at 
low $l$ values according to some 
authors \nocite{Naselsky03a,Naselsky03b}({Naselsky} {et~al.} 2003, 2004). Moreover, even if individual 
pixels in CMB data, rather than statistical properties,
were perfectly resolved into their respective cosmological components, 
it would still be useful to have physically independent checks on CMB
results.

In this article, the method of reducing the noise in 
the cosmic crystallography method --- in the case of a catalogue of AGNs ---
is presented in \SSS\ref{s-method}. This involves successive filters,
parametrised by:
\begin{list}{(\roman{enumi})}{\usecounter{enumi}}
\item $\Delta z$ --- a maximum redshift difference criterion to remove 
many of the noise pairs in a pair catalogue
\item $\Delta\theta$, $\npairs$ --- to remove pairs which are least 
likely to form ``spikes'' of topological origin
\item $\Lselec$ --- to remove selection effect spikes which imitate topological
spikes
\end{list}

Results are presented in \SSS\ref{s-res}
and are summarised in \SSS\ref{s-conclu}.
Notation adopted here for local cosmological parameters includes 
$\Omm$ (present matter density), 
$\Omega_\Lambda$ (cosmological constant and/or quintessence constant),
$w$ (quintessence parameter),
and the Hubble constant parametrised 
as $h\equiv H_0/100$km~s$^{-1}$~Mpc$^{-1}$.

\section{The method}
\label{s-method}

\subsection{The cosmic crystallography method and its variations} 

The cosmic crystallography method was originally proposed by \nocite{LaLu95}{Lachi\`eze-Rey} \& {Luminet} (1995) 
and presented in more detail in \nocite{LLL96}{Lehoucq} {et~al.} (1996). In brief, the idea was the
following. If the Universe (more
specifically: the comoving spatial section of the Universe) is
multiply connected, then each image of a given object is linked to
each of the other images of the object by the holonomies of the space.
Without knowing the curvature or the topology we do not know what
these holonomies are, but we know that they are isometries. Therefore,
any pair of images of a single object can be related by an isometry.
At the separation distance corresponding to such an isometry, the
probability of finding two objects separated by that distance should
be higher than in the absence of the isometry. In other words, the
spatial two-point auto-correlation function of a catalogue of objects should
show excess correlations at certain pair separations.

\nocite{LLL96}{Lehoucq} {et~al.} (1996) presented this as a non-normalised auto-correlation function,
simply plotting counts of pairs of objects and 
calling the function a ``pair separation histogram'' (PSH). 
They carried out several simulations starting from
an artificial catalogue that consisted of a moderate number of objects randomly
distributed in a fundamental polyhedron with equal edge lengths. In such an
idealised case, the excess correlations 
that gather pairs linked by holonomies stand out
dramatically above the background distribution of other pairs, and so 
are termed ``spikes''. \nocite{LLL96}{Lehoucq} {et~al.} (1996)
pointed out that when applied to real data (they used a catalogue of galaxy
clusters), many factors start to play a destructive r\^ole, i.e. by smearing
out such sharp spikes, which makes signatures of the possible existence of ghost
images difficult to detect. \nocite{ULL99b}{Uzan} {et~al.} (1999a) went on to apply the PSH 
test to a quasar catalogue and found no topological signal.

An alternative proposal for
collecting ``generator pairs'' (or ``type II pairs''), as in
cosmic crystallography, was the
the search for local isometries, i.e. to group together
``local pairs'' (or ``type I pairs''),
or more generally, local $n$-tuplets \nocite{Rouk96}({Roukema} 1996).
In \nocite{ULL99a}{Uzan} {et~al.} (1999b), the special case of $n$-tuplets with $n=2$, i.e. pairs,
was discussed and extended.

\nocite{ULL99a}{Uzan} {et~al.} (1999b) extended the $2$-tuplet case 
to provide a single statistic, equivalent to the
auto-correlation function of the spatial 
two-point auto-correlation function, and named this the 
``collecting-correlated-pair'' (CCP) statistic, which should
have a high value in the presence of multiple topological imaging in a
population composed of good standard candles, whether space is hyperbolic,
flat, or spherical. Since there should only be
a significant signal for the correct values of the curvature
parameters $\Omm$ (present matter density), 
$\Omega_\Lambda$ (cosmological constant and/or quintessence constant),
and $w$ (quintessence parameter), 
a signal could potentially be
detected as a relative signal without having to calculate its absolute
significance.

Application of the CCP method to a 
quasar catalogue again gave no topological signature. The authors were
limited by computer time in spanning the full parameter space with high
resolution in parameter values, but the principle is certainly valid. 

Another way to apply a similar test is for the case where:
(1) the curvature parameters'
values are known, (2) the Universe is negatively curved, and (3) a
particular hypothesis is made regarding the 3-manifold of space. This
is known as the ``pullback method'' \nocite{FagG99b}({Fagundes} \& {Gausmann} 1999).

While the PSH test is designed to detect topological lensing, it 
has been pointed out by \nocite{Gomero99a}{Gomero} {et~al.} (2002) that 
(1) it is likely to miss some part of the real signal of topologically lensed
pairs, and (2) it would be insufficient to determine {\em which} 3-manifold
we live in:

\begin{list}{(\arabic{enumi})}{\usecounter{enumi}}

\item The spikes of topological origin in single PSH's are only due to
translations of the covering group, whereas correlations due to the
other (non-translational) isometries manifest as small deformations of the
PSH of the universal covering space. Hence, some part of the 
real signal would be missed (if a rotation in the covering space is needed
when matching faces).

\item Distinct Euclidean manifolds that admit the same translations of
their covering space would present the same spike spectrum, despite
being different; therefore, the set of topological spikes in the
PSHs is not sufficient for distinguishing these compact flat
manifolds, making it clear that even if the comoving spatial section of 
Universe is globally flat and observably multiply connected, 
the spike spectrum may not be enough for determining its global shape.

\end{list}

Another way of stating (1) is that among the locally 
homogeneous, flat 3-manifolds, 
the PSH method works for the 3-torus,
because the distance from a source to its nearest image does not depend on
its location, but for all other locally homogeneous, 
flat 3-manifolds, the PSH method detects only
a subgroup of the full holonomy group, i.e. the subgroup of pure
translations \nocite{Weeks03}({Weeks} {et~al.} 2003). 

However, neither of these limitations suggest that the method should not
be used; any signal that appeared to be significant would imply a 
family of hypotheses to be further tested (depending on 
telescope time allocation).

\subsection{Modifications to the cosmic crystallography
method}
\label{s-the-principles}

Although the active phase of a typical galaxy is
believed to be a recurrent phenomenon rather than happening just once
in its lifetime \nocite{HSE2001}(see e.g. {Hatziminaoglou} {et~al.} 2001), the fraction of time
during which a galaxy is in an active phase is likely to be small.
For quasars, for which the comoving space 
number density at $z\sim$\,1---2
is about a thousand times less than that of galaxies, the fraction of
time during which they are active is likely to be similar, about $10^{-3}$.
This makes the chance of seeing both members of a pair of topological
images small, though not zero.
Neglecting the ``ephemeric'' nature of quasars is the factor that may have
led 
\nocite{ULL99a, ULL99b}{Uzan} {et~al.} (1999b, 1999a)
to have missed a true signal in the catalogue they analysed.

\subsubsection{Maximum redshift difference: $\Delta z$} \label{s-dzfilter}

The short lifetime of AGNs implies that any pair of topological images 
of a single AGN is most likely to be seen during a single period of 
activity of the AGN, i.e. during a short time or redshift interval,
$\Delta z$. From the light travel time argument and assuming that the length of
an activity period is of the order of $10^7$\,years, we adopted
$\Delta z / z =0.005$ as a reasonable amount for a wide range of redshifts.

By removing all pairs of objects at $z_i, z_j$ for which
${|z_i - z_j| / z_i }> {\Delta z / z}$,
the number of potential topological image pairs in the catalogue will be
unchanged, while the total number of pairs will decrease by a large
factor. Hence, the signal-to-noise ratio should increase noticeably.

\subsubsection{Bunches of pairs (BoPs): minimum number of pairs $\npairs$ 
within maximum angular difference $\Delta\theta$} 
\label{s-BoPfilter}

Moreover, if the pairs are due to translations in Euclidean comoving space,
then the vectors drawn between the two members of any pair should be
parallel to one another. Therefore, a second filter can be applied, namely,
selecting those sets of pairs where at least 
$\npairs$ vectors connecting those pairs
point in (nearly) the same direction. Such a collection is referred to as a
`bunch of pairs' (BoP). This means that any {\em lone} 
pair of objects pointing to a given direction, 
isolated from other pairs but fulfilling the
redshift similarity criterion, 
is removed by this filter, so it is not
taken into account at all. 

In this analy\-sis, $\npairs=3$ is adopted; i.e.
if in a given 
direction (within some angular tolerance --- see below)
there are only 
two pairs of objects fulfilling the redshift similarity criterion and have 
vectors in this direction, then these pairs 
are treated as 
coincidences and disregarded. 
In practice, it turned out that the adoption of $\npairs=3$
is very conservative, since even BoPs with $\npairs=5$ constitute
``noise''.

According to this definition, therefore, a BoP consists of at least 3~vectors
connecting pairs of AGNs, and these vectors must fulfill the condition that
the angle between any two of them is $<\Delta\theta$, where $\Delta\theta$
is a small angle.

A physically reasonable value of $\Delta\theta$ depends on possible
errors in the three-dimensional spatial positions of the two AGNs contributing
to a pair. As mentioned in \nocite{Rouk96}{Roukema} (1996), an upper limit of a few
100~km/s on peculiar velocities limits errors in positions to about 
1-10~{\hMpc}. For pair separations of about 3000-6000~{\hMpc}, this
implies maximum errors of about $0.003$ radians.
However, to be conservative, in order not to remove too many pairs that
might correspond to a true signal, a much larger
tolerance can be adopted.

In general, particularly given the non-uniform
distribution of the objects with known redshifts in the catalogue we used,
varying $\Delta\theta$ has a clear effect on the distribution of BoPs.
If a high value of $\Delta\theta$ is set, several BoPs are
merged together, or ``smeared'', while for lower values,
both the ``noise'' and a potential signal are
suppressed.  The smearing effect can be monitored by 
checking the distribution of the angles of vectors within BoPs.
For the present study,
a $\Delta\theta$ which is low enough to avoid the ``smearing'' effect of
BoPs was found by starting with a high value and decreasing 
it until the ``poorly collimated'' BoPs vanished. This value,
$\Delta\theta =0.075$~rad,
is (conservatively) well
above the limit to which angular errors are likely to occur due to
observational effects such as peculiar velocity and the precision of
redshift estimates.

\fdNdz

\subsubsection{Selection effects and how to avoid them: $\Lselec$} 
\label{s-seffects}

If a topological signal is present in an AGN catalogue, then a PSH
filtered for short lifetimes and confined to the BoPs should show a higher
signal-to-noise spike at the length of one of the generators of the
3-manifold of space than in the absence of these filters. However, 
the presently available AGN
catalogues
generally include deep and narrow surveys in which
many objects are located in a small solid angle or else in a long but
narrow angular strip of the sky.\footnote{We deliberately 
ignore the SDSS for this very reason --- see Sect.~\ref{s-generic}
and~\ref{s-conclu}.} 
It is obvious that a pair of (small)
patches of the sky well-sampled with observations increases the probability
that there will be objects of the same redshift in each patch and so 
would mimic a topological signal in a PSH. Effects of clustering are likely
to generate higher than random numbers of parallel
vectors, of nearly equal length, 
between pairs of objects. This is a selection effect 
that is likely to lead
to non-topological spikes in a PSH.

A third filter is possible to counter this effect.
While spikes resulting from this selection effect 
are caused by objects clustered in small regions
of space, spikes that represent topological pairs should occur for objects
spread throughout space.
So, by plotting the vectors represented in a spike in combination with
the spatial positions of the objects from which they originate, 
it should be possible to judge whether the spike is caused by
a selection effect or whether it is consistent with topological 
lensing. More quantitatively, a minimum spatial separation
$\Lselec$ between the objects generating the pairs in a bunch can
be defined in order to exclude most of the pairs from any BoP 
generated by this selection effect.
Any spike of topological origin should remain after the application of
the $\Lselec$ criterion, while artefactual spikes, caused by this
selection effect, should be removed.

\fdNdmag

\section{Results}

\subsection{The sample of AGNs} \label{s-generic}

We used an early version of the catalogue compiled by
\nocite{FleschHard04}{Flesch} \& {Hardcastle} (2004)\footnote{Available at ftp://ftp.quasars.org.}
which aligns and overlays the year 2000 releases of
the ROSAT, HRI, RASS, PSPC, and WGA X-ray catalogues, the NVSS and FIRST
radio catalogues, the Veron QSO catalogue, the Principal Galaxy Catalogue,
and the SDSS-EDR onto the optical APM and USNO catalogues. The version we
used is dated 18 October 2001. It contains 26533 objects with known
redshifts, but we actually took 15762 with $z>0.1$ for our analysis. 
The redshift and apparent magnitude distributions of this sample of
15762 objects are shown in Figs~\ref{f-dNdz} and \ref{f-dNdmag}. 

The dominant observational uncertainties that are relevant for 
pair separation histogram analysis are the redshifts.
Angular positions on the sky are known for these objects
to better than an arcsecond, which corresponds to an error of 
much less than a megaparsec in the sky plane.

Typical redshift uncertainties are $\pm 0.01$ in the redshift value,
corresponding to measurement errors of 
about 30~{\hMpc} at $z=0.10$ or about 10~{\hMpc} 
at $z=3.00$. Additional uncertainty in the true distance estimates
for a fixed set of local cosmological parameters $(\Omm, \Omega_\Lambda, w,
H_0)$ comes from the unknown peculiar velocities of the objects relative
to the comoving coordinate frame. A peculiar velocity of a few 100 km/s 
would give about 1-10~{\hMpc} additional radial uncertainty in position.

We also
attempted to explore a newer version of that catalogue which includes the
Sloan Digital Sky Survey/Data Release 1 (SDSS/DR1) \nocite{SDSS02}({Schneider} {et~al.} 2002). It turned
out, however, that this introduced a large selection effect,
as discussed in \SSS\ref{s-seffects} which had
to be removed anyway. Consequently, we abstained from trying
any of the more recent but --- at the time of writing --- still incomplete
releases of SDSS.

\subsection{Application of the method} \label{s-res}

\subsubsection{Maximum redshift difference: $\Delta z$} \label{s-dzfilter-results}

As discussed in \SSS\ref{s-the-principles}, 
a tight constraint has to be imposed on $\Delta z$
in order
to remove most pairs that do not represent an AGN in a single active
phase, while also including some tolerance to
reflect observational uncertainty in three-dimensional comoving
spatial positions caused by peculiar velocities and 
inaccuracies in measured AGN redshifts. Fig.~\ref{f-PSHdeltaz} shows the PSH
for all objects we used for our analysis (\SSS\ref{s-generic})
assuming local cosmological parameters $\Omm=0.3, \Omega_\Lambda=0.7, w=-1$
after filtering with ${\Delta z / z} = 0.005$.

No obvious signal is present in this figure; despite removing a large
part of the noise, either no signal is present at all, or any signal
present is still hidden. As the following discussion shows, non-topological signals
{\em are} present but require further filtering to successively
detect and then remove them.

\subsubsection{Bunches of pairs (BoPs): minimum number of pairs $\npairs$ 
within maximum angular difference $\Delta\theta$} 
\label{s-BoPfilter-results}

As described in \SSS\ref{s-BoPfilter}, pairs caused by topological generators
should be parallel to each other. 
Fig.~\ref{f-spectr} shows the result of following the 
${\Delta z / z} = 0.005$ criterion by looking for BoPs with
$\Delta\theta = 0.075$~rad and $\npairs=3$.

\fPSHdeltaz
\fspectr
\fpolar 
\fclustp
\fLselec
\ffinal

As can been seen in Fig.~\ref{f-spectr} the concept of BoPs works quite
well as a filter of the ``noise'' introduced by pairs of non-topological
origin: several spikes with good signal-to-noise ratios clearly become
visible when in addition to taking into account 
AGNs' short lifetimes --- 
an effect of the $\Delta z$ filter --- 
we also reject all the pairs that do not constitute BoPs with at least three
members ($\npairs=3$). 

While the signal-to-noise has again been improved, this is still insufficient
to establish the topological nature of the signal.
In particular, the apparently good signal may be the
result of selection effects. Clearly, at least some of those spikes are
likely to be non-topological in origin, since there is a distribution of
many spikes close together, especially for the spikes at around
$2-3${\hGpc}, whereas the error in distances for a true spike should not be
larger than about 10{\hMpc}. If any of the spikes {\em are} topological,
then they just happen to have a signal of about the same amplitude as noise
(or selection effect) spikes. 

\subsubsection{Selection effects and how to avoid them: $\Lselec$} 
\label{s-seffects-results}

Before applying the $\Lselec$ criterion to remove selection effect spikes 
(\SSS\ref{s-seffects}), it is interesting to look at the orientation of all
the BoPs of Fig.~\ref{f-spectr}, which is plotted in Fig.~\ref{f-polar}.
Since the peaks happen to emerge in two groups ($2-3${\hGpc} and
$6-8${\hGpc}), two plots are presented, one for each of these groups. It
turns out that the bifurcation we observe in Fig.~\ref{f-spectr} reflects
the bias caused by the Galaxy, i.e. the scarcity of extragalactic
observations near the galactic equator (the so-called zone of avoidance).
The BoPs present in the upper panel of Fig.~\ref{f-polar}, 
where the distances in pairs are $< 5${\hGpc} 
(the ``short'' ones), connect objects mostly on the same side of
the galactic equator, either north or south. 
Therefore, the ``short'' BoPs predominantly point towards low
galactic latitudes. 

On the other hand, the BoPs present in the lower panel connect the objects
on both sides of the galactic equator, and so they largely point to high
galactic latitudes. This explains why the histogram is bimodally
distributed, but the question of whether any spikes are topological remains.

The most outstanding spikes in Fig.~\ref{f-spectr} are the two 
at about 7000{\hMpc}. However, as mentioned in
\SSS\ref{s-seffects}, what we could and should expect is an
ubiquity of false signals resulting from an uneven distribution of
objects with known redshifts, since spectroscopic observations are
often concentrated in small solid angles.

A visual way to investigate this 
for a particular peak in the histogram of BoPs is to plot the distribution
of pairs in a given BoP in the 
cartesian coordinate system where the $z$ axis
points to the averaged direction that the BoP points to. If a given BoP is a 
result of multiple connectedness, a global property of the 
Universe, then the 
distribution of pairs should not favour any particular area on
the ``equatorial plane'' of this coordinate system; in other words,
the isomorphism from one copy of the fundamental domain to 
another applies equally well everywhere within the fundamental 
domain --- it is a Clifford transformation. (This is by assumption --- 
``generator pairs'', or ``type II pairs'', only give spikes in PSHs if
Clifford transformations are present.)

The spatial distribution of pairs in the BoPs responsible for the two peaks in
Fig.~\ref{f-spectr} around 7000{\hMpc} are shown in Fig.~\ref{f-clustp}.
Clearly, most of these nearly parallel pair vectors
come from a selection effect in the compilation of catalogues we used.
The plots for the other high spikes in Fig.~\ref{f-spectr} are very similar.

As described in \SSS\ref{s-seffects}, this type of selection effect
should be removable by applying the $\Lselec$ criterion. We tried several
values for this parameter, and found that the $\Lselec$ criterion works
effectively for values $\Lselec \ge 150${\hMpc}. By applying 
$\Lselec=150${\hMpc}, 
the ``clumped'' pairs in the BoPs represented in Fig.~\ref{f-spectr}
disappear and, as a result, the $\Lselec$-filtered BoPs 
contain much less pairs. The histogram is now the one shown
in Fig.~\ref{f-PSHLselec}; clearly the BoPs shown in 
Fig.~\ref{f-clustp} no longer contain many more pairs than any other BoPs,
of which most must be BoPs representing noise.

There is no longer any obviously significant signal among the remaining BoPs.
The highest peak
represents the most abundant BoP, with 8 pairs. 
Fig.~\ref{f-final} shows the spatial distribution of the pairs for this
particular BoP. The averaged direction they point in has the following
galactic coordinates: $l=241.4\degr, b=78.2\degr$. The average length of
the vectors in this BoP (i.e. the length of the putative topology generator)
is 6511{\hMpc}.

\section{Discussion, conclusions and future work} \label{s-conclu}

Several filters that remove noise from the pair separation histogram
(PSH) have been presented and applied: these clearly show that the
PSH method can be improved on to find weak signals and to remove false
signals.

The remaining best peak in the filtered PSH, from 
the full AGN compilation, is somewhat similar in its
low significance level to that of \nocite{Weath03}{Weatherley} {et~al.} (2003). 
They repeated the search made by \nocite{FaWi87}{Fagundes} \& {Wichoski} (1987);
i.e. assuming that the Milky Way was once a quasar, they looked for its
``ghost'' images using up-to-date quasar catalogues including the 2dF Galaxy
Redshift Survey (2dFGRS)\footnote{The 2dFGRS is now integrated with the 2dF
QSO survey \nocite{Coll03}({Colless} {et~al.} 2003).} \nocite{Coll01,Coll03}({Colless} {et~al.} 2001, 2003) and SDSS/DR1 \nocite{SDSS02}({Schneider} {et~al.} 2002).
Out of the 7~antipodal
pairs that might be potential topological images of the Galaxy, 6~involve
objects belonging either to 2dFGRS or to SDSS or both. 
So, removal  
of the 2dFGRS and SDSS data from the analysis by \nocite{Weath03}{Weatherley} {et~al.} (2003) 
would yield only one solution. As in our own analysis, an apparent
7-fold increase is capable of mimicking a strong topological signal;
the authors conclude that their result shows no significant
difference in the number of pairs found over the number expected purely by
chance.

It seems there are two ways to correct for this inhomogeneity. The first one
is obvious: one has to use a comprehensive, large area (whole sky, at
best) catalogue obtained via a uniform procedure. The SDSS, when completed,
will be a very
good resource for tackling this problem, even though it will 
only cover $\pi$ steradians\footnote{See http://www.sdss.org for details.}. 

Note, however, that the early SDSS releases, especially SDSS/DR1 
\nocite{SDSS02}({Schneider} {et~al.} 2002)\footnote{http://www.sdss.org/dr1/} have an intrinsic problem
that outweighs any of the normal observational problems such as 
apparent magnitude incompleteness. Since 
these observations are carried out in relatively narrow strips 
($\sim 5-10\degr$), of which two of the main ones lie 
along opposite sides of the celestial equator --- and one lies fairly near the
celestial equator --- a large fraction of the quasars in the catalogue 
necessarily lie
in what is nearly a single plane. So, many separation vectors will
necessarily be nearly parallel and have nearly identical lengths, because
that is the geometry of the catalogue. 

As the SDSS increases to cover a large solid angle, this problem will
become less important but will perhaps persist. Thus, the filters presented
in this paper, $\Delta \theta$ in \SSS\ref{s-BoPfilter} and $\Lselec$ in
\SSS\ref{s-seffects}, still will be needed. They could
potentially be ``tuned'' to the specific geometry of the SDSS in order to
filter out the expected ``selection effect'' spikes caused by the survey
geometry. Additional filters with spatial orientations specifically 
designed for the celestial positions of the SDSS selection function, 
in contrast to the generic filters presented here,
could also help remove false signals.

It is interesting to note at this point that the 
SDSS has its radio counterpart:
the {\it VLA Faint Images of the Radio Sky at Twenty-cm survey}
\nocite{FIRST}(FIRST, {White} {et~al.} 1997); and this FIRST catalogue 
might be used
instead of SDSS. 
The radio oriented approach has an important virtue
that might help not only to detect a 
possible topological signal in the future
when the complete SDSS data becomes available, but also 
help to further test a candidate generator and to choose the
right topological model, if such a signal exists. Instead of considering AGNs
as mere point-like objects, as has always been done so far, one can
explore their morphologies. The FIRST catalogue contains 811,000 sources,
which implies that at the apparent magnitude 
$m(v)\sim 24$ limit of SDSS, $\sim 50$\,\% of the optical
counterparts to FIRST sources will be detected. 

The availability of the {\em maps} (two-dimensional images)
of radio objects listed in the FIRST catalogue provides a
completely new possibility to design a statistical test, such as the
PSH method combined with the filters presented above, by 
comparison of the morphologies of potential pairs
of topologically lensed radio-loud AGNs (RLAGNs).

We do realise that the power of this method is limited by a crucial factor:
radio structures of AGNs are known to evolve rather quickly. This requires 
fine accuracy in the redshift data. 
When fully available, the SDSS, with
an accuracy of radial velocity estimation reaching 30\,km~$\mathrm s^{-1}$,
will give an opportunity to apply this method. In other words, if there is a
statistically significant clue that a set of objects seen at identical
redshifts could be pairs of topologically lensed images, then
they should exhibit
a substantial degree of morphological similarity, after correction 
for the projection effects caused by different viewing angles. 

These projection effects imply another caveat: the 
beaming effect changes the
appearance of RLAGNs considerably, so the search for topologically lensed
images has to be confined to objects in nearly antipodal positions within a
pair. If that constraint is pushed even further, i.e. to (almost) exact
antipodal positions, then this method becomes restricted to a special case,
i.e. to the approach explored by
\nocite{FaWi87}{Fagundes} \& {Wichoski} (1987) and followed-up by \nocite{Weath03}{Weatherley} {et~al.} (2003),
and provides yet
another chance to check if we can see topological ghosts of the Milky Way.
The difference between the two methods in that case is that for 
an RLAGN pair at a redshift $z$ corresponding to distance $L$ from
the observer, \nocite{FaWi87}{Fagundes} \& {Wichoski} (1987) and \nocite{Weath03}{Weatherley} {et~al.} (2003) only consider 
generators of length $L$, while the more general PSH method, constrained
to (almost) exact antipodal positions, considers both generators of 
length $L$ and of length $2L$.

Our results can be summarised as follows:
\begin{list}{(\arabic{enumi})}{\usecounter{enumi}}
\item 
Application of the $\Delta z/z = 0.005$ filter
(\SSS\ref{s-dzfilter}, \SSS\ref{s-dzfilter-results})
to a catalogue of 
15762 objects with $z > 0.1$ 
from an observational compilation of AGN catalogues 
enables removal of most of the 
roughly $10^8$ pairs from the pair separation histogram (PSH), which 
for physical reasons --- the short lifetimes of AGNs --- 
are much more likely to be noise than pairs 
satisfying the filter. This is shown in Fig.~\ref{f-PSHdeltaz}.
\item Application of the $\Delta \theta =0.075$~rad and $\npairs=3$ filters,  
which select out only those 
bunches of pairs (BoPs) which are consistent with being caused by
an isomorphism (generator) between copies of the fundamental polyhedron
(\SSS\ref{s-BoPfilter}, \SSS\ref{s-BoPfilter-results}),
and thereby further
removes
noise from the PSH,
results in further improvement in the signal-to-noise, 
and yields many BoPs which  
become obviously visible --- see Fig.~\ref{f-spectr}. 
\item 
However, these BoPs are best interpreted as selection effect 
spikes (\SSS\ref{s-seffects}). 
A length criterion $\Lselec150$~{\hMpc} 
to remove those BoPs most likely to be selection effects is 
effective (\SSS\ref{s-seffects-results}) 
in doing so.
\item 
The highest spike remaining after these filters (Fig.~\ref{f-PSHLselec}),
points to the galactic direction ($l=241.4\degr, b=78.2\degr$) and
is of length 6511{\hMpc}. However, it is not highly significant.
\end{list}

When the full SDSS becomes available, applying this
method would definitely be worth trying in order to extract any 
signal that might be present. Even though WMAP analyses have so
far indicated that no topological signal is likely to be present
at scales much below the horizon scale, it would be imprudent to
rely on one observational method alone to consider such a fundamental
cosmological property --- the shape of the Universe --- 
to be unmeasurable on these scales.

Another prospect for future work is that 
if an interesting signal were found, then comparing radio 
maps of the putative topologically lensed RLAGNs would 
seem to be a promising tool for
discriminating between true and false topological signals.

\begin{acknowledgements}

SB acknowledges support from KBN Grant 1P03D 012 26.

\end{acknowledgements}

\subm{\clearpage}

\nice{
%

}

\newcommand\myoldbib{


}

\end{document}